\documentclass[12pt]{article}
\usepackage{amsmath,amsfonts,amsthm,amssymb,mathrsfs}
\usepackage{times}
\usepackage{color,ulem}
\usepackage{appendix}
\theoremstyle{plain}
\newtheorem{theorem}{Theorem}[section]
\newtheorem{lemma}[theorem]{Lemma}

\newtheorem{cor}[theorem]{Corollary}
\theoremstyle{definition}

\theoremstyle{remark}
\newtheorem*{remark}{Remark}

\DeclareMathOperator{\sgn}{sgn}

\newcommand{\R}{\mathbb{R}}
\newcommand{\F}{\mathbb{F}}

\renewcommand{\tilde}{\widetilde}
\renewcommand{\hat}{\widehat}

\newcommand{\RN}[1]{%
  \textup{\uppercase\expandafter{\romannumeral#1}}%
}
\begin{document}

\title{Near horizon limit of the Wang--Yau quasi-local mass}
\author{Po-Ning Chen}
\date{}

\maketitle

\begin{abstract}

In this article, we compute the limit of the Wang--Yau quasi-local mass on a family of surfaces approaching the apparent horizon (the near horizon limit). Recently, Pook-Kolb, Zhao, Andersson, Krishnan, and Yau investigated the near horizon limit of the Wang--Yau quasi-local mass in binary black hole mergers in \cite{PK-Z-A-K-Y} and conjectured that the optimal embeddings approach the isometric embedding of the horizon into $\R^3$. Moreover, the quasi-local mass converges to the total mean curvature of the horizon in $\R^3$. The vanishing of the norm of the mean curvature vector implies special properties for the Wang--Yau quasi-local energy and the optimal embedding equation. We utilize these features to prove the existence and uniqueness of the optimal embedding and investigate the minimization of the Wang--Yau quasi-local energy. In particular, we prove the continuity of the quasi-local mass in the near horizon limit.
\end{abstract}

\section{Introduction}\label{sec1}

Given a spacelike surface $\Sigma$ in a spacetime $N$, the Wang--Yau quasi-local energy $E(\Sigma, X, T_0)$ is defined in \cite{W-Y1,W-Y} with respect to each pair $(X, T_0)$ of an isometric embedding $X$ of $\Sigma$ into the Minkowski space $\R^{3,1}$ and a constant future timelike unit vector $T_0\in \R^{3,1}$. The data used in the definition of the Wang--Yau quasi-local energy is the triple $(\sigma,|H|,\alpha_H)$ on $\Sigma$ where $\sigma$ is the induced metric, $|H|>0$ is the norm of the mean curvature vector, and $\alpha_H$ is the connection 1-form (see \eqref{connection-form}). If the spacetime satisfies the dominant energy condition, then $E(\Sigma, X, T_0) \ge 0$ for any admissible pair $(X,T_0)$ (see \cite[Definition 5.1,Theorem A]{W-Y}. The inner product
\[
\tau= - X \cdot T_0
\]
is the time function on $\Sigma$. We also denote the energy as $E(\Sigma,\tau)$ since $E(\Sigma, X, T_0)$ depends only on $\tau$ instead of the pair $(X,T_0)$.

The Wang--Yau quasi-local mass is defined to be the infimum of the quasi-local energy $E(\Sigma,\tau)$ among all admissible time functions $\tau$. The Euler--Lagrange equation for the critical points of the quasi-local energy is derived in \cite{W-Y}. This equation is called the optimal embedding equation, and the solution is referred to as the optimal embedding since it identifies the best configuration of $\Sigma$ in the reference Minkowski space. This configuration serves as the ground state for the gravitational energy.

The optimal embedding equation is a fourth-order partial differential equation (Equation \eqref{optimal-W-Y}) coupled with the isometric embedding equation for the surface into the Euclidean space $\R^3$. The optimal embedding equation plays a fundamental role in the definition of the Wang--Yau quasi-local mass and its application in general relativity. In the generic situation, the equation is highly challenging to solve. However, if one considers a family of surfaces in the spacetime whose induced data converges to that of a surface in Minkowski space (weak field limit), the optimal embedding equation can be solved using perturbative methods. This approach is used in the large sphere limit of the Wang--Yau quasi-local mass at both spatial infinity and null infinity and the small sphere limit near a point \cite{W-Y3,C-W-Y1,C-W-Y3}. In particular, the solution of the optimal embedding equation near the null infinity plays a fundamental role in the supertranslation invariance of CWY angular momentum at null infinity \cite{K-W-Y,CKWWY,CWWY}.

Another key observation is that $\tau=0$ solves the optimal embedding equation if
\[
div_{\sigma} \alpha_H = 0.
\]
Miao, Tam, and Xie solved the optimal embedding equation using the implicit function theorem assuming $div_{\sigma} \alpha_H$ is sufficiently small in \cite{M-T-X}. This result is closely related to the second variation of the Wang--Yau quasi-local energy \cite{C-W-Y2,M-T}.

Similar to  \cite{W-Y3,C-W-Y1,C-W-Y3}, it is important to solve the optimal embedding equation in order to evaluate the Wang--Yau quasi-local mass in the near horizon limit. For surfaces in the static slice of the  Schwarzschild spacetime, $\tau=0$ solves the optimal embedding equation, and the Wang--Yau energy for $\tau=0$ reduces to the Brown--York mass. The near horizon limit yields a value of twice the ADM mass. See \cite{B-Y,Martinazzi,D-T} for the near horizon limit of various quasi-local masses in the Kerr spacetime. However, it is less clear how to solve the optimal embedding equation for non-stationary black holes. 

An iterative scheme based on the method used in \cite{C-W-Y1} was proposed in  \cite{Chen} to solve the optimal embedding equation for surfaces near an apparent horizon. In particular, the results in \cite{Chen} suggested that there exists a family of optimal embeddings that converge to the isometric embedding of the horizon into $\R^3$ as the surface approaches the horizon. 

In \cite{PK-Z-A-K-Y}, Pook-Kolb, Zhao, Andersson, Krishnan, and Yau studied the Wang--Yau quasi-local energy in binary black hole mergers. In particular, it is conjectured that for a family of surfaces approaching the apparent horizon, there is a unique family of solutions to the optimal embedding equation near $\tau=c$ which realizes the Wang--Yau quasi-local mass. By the translation invariance of the quasi-local mass, it suffices to consider $\tau=0$. Accordingly, we consider time functions whose average on $\Sigma$ are zero. The mean curvature vector is null on the horizon. In particular, the conjecture includes a proposal to define Wang--Yau quasi-local mass for surfaces with null mean curvature vector. In \cite{Z-A-Y}, Zhao, Andersson, and Yau further investigated the behavior of the Wang--Yau quasi-local energy and the optimal embedding equation in the near horizon limit and proved that the optimal embedding equation admits no solution near the horizon if the horizon can't be isometrically embedded into $\R^3$ and study the limit of the quasi-local mass if the horizon admit positive Gauss curvature.

In this article, we first defined the quasi-local mass $ m_{WY} (\Sigma_{o})$ on the apparent horizon $\Sigma_{o}$ following the min-max principle of the Wang--Yau quasi-local mass \cite{W-Y} for surfaces with spacelike mean curvature vectors. In particular, on the horizon, $\tau=0$ is the unique global minimizer of the Wang--Yau quasi-local energy, and the value of the Wang--Yau quasi-local mass agrees with \cite[Remark VI.5]{PK-Z-A-K-Y}.

Next, we consider a family of surfaces $\Sigma_s$ approaching the apparent horizon $\Sigma_{o}$ as $s$ approaches $0$ \footnote{We will address the necessary regularity for each result separately. However, one may simply consider a smooth family converging smoothly to the horizon. }. We assume that the Gauss curvature $K_s$ of $\Sigma_s$ is positive and uniformly bounded away from zero. Moreover, we assume that the mean curvature vector $H_s$ for $\Sigma_s$ is spacelike for each $s>0$. We prove that there is a family of solutions $\tau_s$ to the optimal embedding equation on $\Sigma_s$ for $s$ sufficiently small. Moreover,
\[
\lim_{s \to 0} \tau_s = 0
\]
and
\begin{equation}\label{continuity_WY}
\lim_{s \to 0} m_{WY}(\Sigma_s) = \lim_{s \to 0} E_{WY}(\Sigma_s,\tau_s) = m_{WY} (\Sigma_{o}).
\end{equation}
This property is what we refer to as the continuity of the Wang--Yau quasi-local mass in the near horizon limit.

To investigate the near horizon limit, we apply results and methods for minimizing the Wang--Yau quasi-local energy from \cite{C-W-Y1,C-W-Y2,M-T-X,M-T,W-Y}. In particular, we will use the elliptic estimate of isometric embedding established by Nirenberg \cite{Nirenberg}, which is used in \cite{M-T-X}. There are several advantages when one considers a family of surfaces approaching the apparent horizon:

\begin{enumerate}
\item One can iterate the optimal embedding equation using the contraction mapping theorem, whereas, in the previous settings, one typically uses the implicit function theorem.
\item The vanishing of the physical mean curvature suggests the convexity of the Wang--Yau quasi-local energy.
\item For any given time function $\tau$, the quasi-local energy $E(\Sigma_s,\tau)$ approaches positive infinity as $\Sigma_s$ approaches the horizon. 
\end{enumerate}
We will use the first property to establish the existence of the solution to the optimal embedding equation. We will then use the third property to study the near horizon limit of the Wang--Yau quasi-local mass in the appendix. We will further apply these properties to address the uniqueness and minimizing properties of the solution to the optimal embedding equation. See also \cite{Z-A-Y} for a further analysis of the third point.

In Section \ref{sec2}, we review the definition of Wang--Yau quasi-local mass for surfaces with spacelike mean curvature. In Section \ref{sec3}, we utilize the min-max principle to define the Wang--Yau quasi-local mass for surfaces with null mean curvature. In Section \ref{sec4}, we solve the optimal embedding equation in the near horizon limit using the contraction mapping theorem. In Section \ref{sec5}, we prove the continuity of the Wang--Yau quasi-local mass in the near horizon limit. In the appendix, we identify the leading-order term of the solution found in Section \ref{sec4} and compare it to the results from \cite{Chen}. We will also discuss the convexity of the Wang--Yau quasi-local energy in the near horizon limit.

\subsection{Setup for the near horizon limit}
Throughout our discussion in the remaining article, we assume that the horizon $\Sigma_o$ admits positive Gauss curvature.  Moreover, we assume that there is a choice of a spacelike unit normal of $\Sigma_o$ such that \footnote{We assume the same convention for the mean curvature vector as in \cite{W-Y1,W-Y}, thus the choice of negative sign in the equation below.}
\[
-\langle H_o ,e_3\rangle \ge 0
\] 
Such a condition is true if one considers an outward area-minimizing horizon in an asymptotically flat initial data set. However, we would like to keep the condition as local as possible. 

Furthermore, for the near horizon limit, we assume that there is a family of surfaces $\Sigma_s$ that converges (smoothly or at least sufficiently regularly) to $\Sigma_o$, as $s$ goes to $0$. For each $s>0$, the mean curvature $H_s$ of $\Sigma_s$ is spacelike, and the Gauss curvature of $\Sigma_s$ is positive. The metric $\sigma_s$ converges to the induced metric on the horizon, and the norm of the mean curvature vector $|H_s|$ converges to 0. The required norms for convergence differ for the various results and will be specified in the later sections. Finally, we assume that 
\[
div_{\sigma_s} \alpha_{H_s}
\]
is uniformly bounded (again in norms to be specified). 

One important application is to study the outer-area minimizing horizon on the asymptotically flat initial data set. In such cases, the normal $e_3$ would point toward the outward direction and the surfaces $\Sigma_s$ foliate a neighborhood of $\Sigma_o$ on the initial data set on the side of $e_3$. See \cite{Z-A-Y} for discussion on the construction of a family of such surfaces in a similar setting (particularly on boundedness of $div_{\sigma_s} \alpha_{H_s}$).

\section{Review of Wang--Yau quasi-local mass}\label{sec2}

In this section, we review the definition of the Wang--Yau quasi-local mass and the optimal embedding equation from \cite{W-Y1,W-Y}. We emphasize the min-max principle used to define the quasi-local mass since we will follow the same principle to define the Wang--Yau quasi-local mass on the horizon in Section \ref{sec3}.

Let $\Sigma$ be a closed embedded spacelike 2-surface in a spacetime $N$ with spacelike mean curvature vector $H$. Let $J$ be the reflection of $H$ through the future outgoing light cone in the normal bundle of $\Sigma$.

The data used in the definition of the Wang--Yau quasi-local energy is the triple $(\sigma,|H|,\alpha_H)$ on $\Sigma$ where $\sigma$ is the induced metric, $|H|$ is the norm of the mean curvature vector, and $\alpha_H$ is the connection 1-form of the normal bundle with respect to the mean curvature vector
\begin{equation} \label{connection-form}
\alpha_H(\cdot )=\langle \nabla^N_{(\cdot)} \frac{J}{|H|}, \frac{H}{|H|} \rangle \end{equation}
where $\nabla^N$ is the covariant derivative in $N$.

Given an isometric embedding $X:\Sigma\rightarrow \R^{3,1}$ and a future timelike unit vector $T_0$ in $\R^{3,1}$, suppose the projection $\widehat{X}$ of $X(\Sigma)$ onto the orthogonal complement of $T_0$ is embedded, and denote the induced metric, the second fundamental form, and the mean curvature of the image surface $\widehat{\Sigma}$ of $\widehat{X}$ by $\hat{\sigma}_{ab}$, $\hat{h}_{ab}$, and $\widehat{H}$, respectively.

The Wang--Yau quasi-local energy $E(\Sigma, X, T_0)$ of $\Sigma$ with respect to the pair $(X, T_0)$ is defined to be the maximum of the difference of the total mean curvature and total generalized mean curvature
\[
\frac{1}{8\pi}\int_{\hat \Sigma} \widehat H d \widehat \Sigma- \frac{1}{8\pi} \int - \sqrt{1+|\nabla \tau|^2} \langle H , e_3 \rangle - \alpha_{e_3}(\nabla \tau) d \Sigma
\]
among all frame $\{e_3,e_4 \}$ of the normal bundle over $\Sigma$ where $\nabla$ and $\Delta$ are the gradient and Laplace operator of $\sigma$, respectively, and $\tau=-\langle X, T_0\rangle$.

From \cite[Proposition 2.1]{W-Y}, the  minimum of the total generalized mean curvature (hence the maximum of the difference) is achieved by the frame with angle
\[\theta=\sinh^{-1}(\frac{-\Delta\tau}{|H|\sqrt{1+|\nabla\tau|^2}}),\]
to the frame determined by the mean curvature vector \footnote{It worths noting that the spacelike mean curvature condition is key in proving \cite[Proposition 2.1]{W-Y}. This is the main reason why this condition is needed for the Wang--Yau energy.}. The Wang--Yau quasi-local energy is
\[
\begin{split}E(\Sigma, X, T_0)
= \frac{1}{8\pi}\int_{\widehat{\Sigma}} \widehat{H} d{\widehat{\Sigma}} - \frac{1}{8\pi}\int_\Sigma \left[\sqrt{1+|\nabla\tau|^2}\cosh\theta|{H}| - \nabla \tau \cdot \nabla \theta -\alpha_H ( \nabla \tau) \right]d\Sigma.\end{split}\]
The Wang--Yau quasi-local mass is defined to be the infimum of the Wang--Yau quasi-local energy $E(\Sigma, X, T_0)$ among all admissible pairs $(X,T_0)$.

Finally, we recall that the Euler--Lagrange equation for a critical point $(X_0, T_0)$ of the quasi-local energy is
\begin{equation} \label{optimal-W-Y}
-(\widehat{H}\hat{\sigma}^{ab} -\hat{\sigma}^{ac} \hat{\sigma}^{bd} \hat{h}_{cd})\frac{\nabla_b\nabla_a \tau}{\sqrt{1+|\nabla\tau|^2}}+ div_\sigma (\frac{\nabla\tau}{\sqrt{1+|\nabla\tau|^2}} \cosh\theta|{H}|-\nabla\theta-\alpha_{H})=0
\end{equation}
coupled with the isometric embedding equation of $\hat \sigma = \sigma_{ab}+  d \tau \otimes  d \tau$ into $\R^3$.

\section{Wang--Yau quasi-local mass on the apparent horizon}\label{sec3}
In this section, we use the same min-max principle described in Section \ref{sec2} to define quasi-local mass on the horizon $\Sigma_o$ where the mean curvature vector is null.

Fixing the reference Hamiltonian to be
\[
\int_{\hat \Sigma} \widehat H d \widehat \Sigma,
\]
we focus on the total generalized mean curvature
\[
\int_{\Sigma_o} - \sqrt{1+|\nabla \tau|^2} \langle H , e_3 \rangle - \alpha_{e_3}(\nabla \tau) d \Sigma_o
\]
to derive the physical Hamiltonian for the pair $(X,T_0)$. We have the following lemma concerning the total generalized mean curvature:
\begin{lemma} \label{min_general_horizon}
Consider an apparent horizon $\Sigma_o$ with spacelike unit normal $e_3$ such that 
\[
p= - \langle H , \breve e_3 \rangle \ge 0.
\]
Then, for any time function $\tau$ which is not identically a constant, the infimum of the total generalized mean curvature
\[
\int_{\Sigma_o} - \sqrt{1+|\nabla \tau|^2} \langle H , e_3 \rangle - \alpha_{e_3}(\nabla \tau) d \Sigma_o
\]
among all frame $\{e_3,e_4 \}$ of the normal bundle over $\Sigma_o$ is negative infinity. On the other hand, for $\tau=c$, the infimum is zero
\end{lemma}

\begin{proof}
In the following proof, all the integrals are computed on $\Sigma_o$ with the induced metric. We choose any base frame $\{\breve e_3, \breve e_4 \}$ and write any other frame as
\[
e_3 = \cosh \phi \breve e_3 + \sinh \phi \breve e_4.
\]
We get
\[
\int - \sqrt{1+|\nabla \tau|^2} \langle H , e_3 \rangle - \alpha_{e_3}(\nabla \tau) = \int \sqrt{1+|\nabla \tau|^2} p (\cosh \phi -\sinh \phi) + \phi \Delta \tau - \alpha_{\breve e_3}(\nabla \tau)
\]
If $\tau$ is not constant, consider
\[
\phi = C_1 - C_2 \Delta \tau.
\]
For positive constants $C_1 \gg C_2 \gg 1$, $\phi$ goes to infinity and $\cosh \phi -\sinh \phi$ goes to zero. The last term is independent of $\phi$, whereas the second term goes to negative infinity.

On the other hand, if $\tau =c$, then the second and the third term vanish, whereas the first term is non-negative and goes to zero as $\phi$ goes to infinity since $p \ge 0$.
\end{proof}

\begin{cor} \label{mass_on_horizon}
Assuming  the same condition as Lemma \ref{min_general_horizon} and that the metric on $\Sigma_o$ admits positive Gauss curvature, then we have
\[
m_{WY} (\Sigma_o )= \frac{1}{8 \pi} \int_{\Sigma_o} H_0
\]
where $H_0$ is the mean curvature of the image of isometric embedding of $\Sigma_o$ into $\R^3$
\end{cor}

\begin{proof}
From Lemma \ref{min_general_horizon}, the Wang--Yau quasi-local energy with respect to the time function $\tau=c$ is precisely $\frac{1}{8 \pi} \int_{\Sigma_o} H_0$. On the other hand, the quasi-local energy is positive infinity for any non-constant time function $\tau$.
\end{proof}

\section{Contraction mapping and existence of optimal embedding}\label{sec4}
To solve the optimal embedding equation, we set up an iteration and use the smallness of $|H|$ to show that this iteration map is a contraction mapping. Recall that the optimal embedding equation is:
\[
-(\widehat{H}\hat{\sigma}^{ab} -\hat{\sigma}^{ac} \hat{\sigma}^{bd} \hat{h}_{cd})\frac{\nabla_b\nabla_a \tau}{\sqrt{1+|\nabla\tau|^2}}+ div_\Sigma (\frac{\nabla\tau}{\sqrt{1+|\nabla\tau|^2}} \cosh\theta|{H}|-\nabla\theta-\alpha_H)=0
\]
with $\sinh \theta =\frac{-\Delta \tau}{|{H}|\sqrt{1+|\nabla \tau|^2}}$.

Consider the following map:
\[
\mathbb F: C^{5,\alpha} \to C^{5,\alpha}
\]
where
\[
\F(\tau) =\tilde \tau
\]
is defined as follows: We first consider the equation 
\begin{equation} \label{iteration-1}
\Delta \tilde \theta = -(\widehat{H}\hat{\sigma}^{ab} -\hat{\sigma}^{ac} \hat{\sigma}^{bd} \hat{h}_{cd})\frac{\nabla_b\nabla_a \tau}{\sqrt{1+|\nabla\tau|^2}}+ div_\Sigma (\frac{\nabla\tau}{\sqrt{1+|\nabla\tau|^2}} \cosh\theta|{H}|-\alpha_H)
\end{equation}
where $\widehat{H}$, $\hat{\sigma}^{ab}$ and $\hat{h}_{cd}$ as well as $\theta$ on the left-hand side come from the isometric embedding with time function $\tau$. The integral of the right-hand side vanishes on $\Sigma$. Indeed, the second term integrates to zero since it is a divergence term. For the first term, recall that (for example, \cite[Section 6]{W-Y})
\[
\begin{split}
d\Sigma =& \frac{d \widehat \Sigma}{\sqrt{1+|\nabla \tau|^2}} \\
\nabla_a\nabla_b \tau =& (1+|\nabla \tau|^2) \hat\nabla_a \hat\nabla_b \tau
\end{split}
\]
and thus
\[
\begin{split}
    &  \int_{\Sigma} (\widehat{H}\hat{\sigma}^{ab} -\hat{\sigma}^{ac} \hat{\sigma}^{bd} \hat{h}_{cd})\frac{\nabla_b\nabla_a \tau}{\sqrt{1+|\nabla\tau|^2}} d \Sigma \\
= & \int_{\Sigma} (\widehat{H}\hat{\sigma}^{ab} -\hat{\sigma}^{ac} \hat{\sigma}^{bd} \hat{h}_{cd}) \hat \nabla_b \hat \nabla_a \tau  d \widehat \Sigma  \\
= &  - \int_{\Sigma} \hat \nabla_b  (\widehat{H}\hat{\sigma}^{ab} -\hat{\sigma}^{ac} \hat{\sigma}^{bd} \hat{h}_{cd}) \hat \nabla_a \tau  d \widehat \Sigma \\
= & 0
\end{split}
\]
where the Codazzi equation of surface $\widehat \Sigma$ in $\R^3$ is used in the last equality. As a result, the solution $ \tilde \theta$ exists and is unique up to an additive constant. Moreover, this is a unique choice of the constant such that the following equation is solvable for $\tilde \tau$
\begin{equation} \label{iteration-2}
\Delta \tilde \tau   = - |{H}|\sqrt{1+|\nabla \tau|^2} \sinh \tilde \theta
\end{equation}
since $\sinh$ is a monotone function. We require the normalization
\[
\int_{\Sigma} \tilde \tau  d\Sigma=0.
\]
A fixed point of the map $\F$ corresponds to a solution to the optimal embedding equation. We first prove the existence of a solution to the optimal embedding equation on a surface $\Sigma$ with a fixed induced metric and connection 1-form when $|H|$ is sufficiently small. Then we apply this result to a family of surfaces approaching the horizon assuming that the physical data $\sigma$ and $\alpha_H$ are controlled and $|H|$ is small in suitable $C^{k,\alpha}$ spaces.

\begin{theorem} \label{optimal-existence}
On a surface $\Sigma$ with a given induced metric $\sigma$ of positive Gauss curvature and a given connection 1-form $\alpha_H$ \footnote{We think of this one form as fixed, independent of the smallness of $|H|$ to be used in the optimal embedding equation.}. There exists $\epsilon$ such that the optimal embedding equation admits a solution provided that
\[
0<\Big{\Vert}  |H| \Big {\Vert} _{C^{3,\alpha}} < \epsilon
\]
\end{theorem}
\begin{proof}

Let $\mathfrak A$ be the subset of $ C^{5,\alpha}$ consisting of functions such that the Gauss curvature of
\[
K(\sigma+ d\tau \otimes d \tau)>0.
\]
By our assumption,
\[
0 \in \mathfrak A
\]
and there exists $c>0$ so that $\tau \in \mathfrak A$ for
\[
\Vert \tau \Vert_{C^{5,\alpha}} < c.
\]

In the following, we restrict $\mathbb  F$ to the subset of functions in $\mathfrak A$ with $\Vert \tau \Vert_{C^{5,\alpha}} < c$ and vanishing average on $\Sigma$. Denote this set by $B_{c}$. We have the following estimate for isometric embedding with different time functions (see \cite[Section 4]{M-T-X}):
\begin{equation} \label{estimate_iso_component}
\Vert X(\tau_1) - X(\tau_2) \Vert_{C^{3,\alpha}} < C \Vert \tau_1 - \tau_2\Vert_{C^{5,\alpha}}
\end{equation}
where $X(\tau)$ for a given function $\tau$ is the isometric embedding of the metric $\sigma+ d\tau \otimes d \tau$ into $\R^3$.

This implies the following estimate on the mean curvature $\widehat{H}$ and second fundamental form $\hat{h}$ with two time functions $\tau_1$ and $\tau_2$:
\[
\begin{split}
\Vert \hat{h}(\tau_1) - \hat{h}(\tau_2) \Vert_{C^{1,\alpha}} < & C \Vert \tau_1 - \tau_2\Vert_{C^{5,\alpha}} \\
\Vert \widehat{H}(\tau_1) - \widehat{H}(\tau_2)\Vert_{C^{1,\alpha}} < & C \Vert \tau_1 - \tau_2\Vert_{C^{5,\alpha}}. \\
\end{split}
\]

We subtract the equation \eqref{iteration-1} with $\tau_1$ and $\tau_2$ to cancel the last term on the right-hand side. Using the above estimate, the difference of the right-hand side in $C^{1,\alpha}$ norm is then bounded by $ \Vert \tau_1 - \tau_2\Vert_{C^{4,\alpha}}$. Standard elliptic estimate shows that (assuming $\tilde \theta_1$ and $\tilde \theta_2$ both average to zero)
\[
\Vert\tilde \theta_1 - \tilde \theta_2\Vert_{C^{3,\alpha}} < C \Vert \tau_1 - \tau_2\Vert_{C^{5,\alpha}}
\]
In terms of $\tilde \tau_1$ and $\tilde \tau_2$, we get
\[
\Delta (\tilde \tau_1 - \tilde \tau_2)= |H| \left ( \sqrt{1+|\tau_2|^2}\sinh (\tilde \theta_2 + D_2) - \sqrt{1+|\tau_1|^2}\sinh (\tilde \theta_1 +D_1) \right)
\]
where $D_1$ and $D_2$ are the unique constants such that
\[
\int |H| \sqrt{1+|\tau_i|^2}\sinh (\tilde \theta_i + D_i) = 0.
\]
We also have that
\[
|D_2 - D_1| < C \Vert \tau_1 - \tau_2\Vert_{C^{5,\alpha}}.
\]
A standard elliptic estimate shows that
\[
\Vert \tilde \tau_1 - \tilde \tau_2 \Vert_{{C^{5,\alpha}} } < C\epsilon \Vert \tau_1 - \tau_2 \Vert_{C^{5,\alpha}}
\]
where we obtain the small constant $\epsilon$ from $\Vert |H| \Vert_{C^{3,\alpha}}< \epsilon $. As a result, for $\epsilon$ sufficiently small, the map is a contraction mapping. Moreover, it is easy to check that
\[
\Vert \F(0)\Vert_{C^{5,\alpha}} < C \epsilon
\]
As a result, for $\epsilon$ sufficiently small, $F$ maps $B_{c}$ to itself. By the contraction mapping theorem, there exists a unique fix point of $F$ in $B_{c}$.
\end{proof}

\begin{remark}
From the contraction mapping theorem, the solution is unique within $B_{c}$.
\end{remark}

\begin{cor} \label{bounded_ratio}
The solution obtained in Theorem \ref{optimal-existence} converges to $0$ in $C^{5,\alpha}$ as $|H|$ converges to $0$ in $C^{3,\alpha}$.
\end{cor}
We apply the above theorem to a family of surfaces $\Sigma_s$ approaching the horizon to find a family of solutions to the optimal embedding equation when $s$ is sufficiently small. To guarantee that the constants appearing in the proof remain uniformly bounded, we require the family of data to be controlled in suitable   $C^{k,\alpha}$ spaces. 

\begin{theorem}  \label{optimal-existence-family}
Assume that the horizon $(\Sigma_o,\sigma_o)$ is outward minimizing with positive Gauss curvature. Suppose we have a family of surfaces $\Sigma_s$ approaching the horizon $\Sigma_o$ such that the family of data $(\sigma_s,|H_s|,\alpha_{H_s})$ satisfies the following: $\sigma_s$ converges to $\sigma_o$ in $C^{5,\alpha}$, $|H_s|$ converges to $0$ in $C^{3,\alpha}$ and $div_{\sigma_s} \alpha_{H_s}$ is uniformly bounded in $C^{3,\alpha}$. Then there exists a family of solution $\tau_s$ to the optimal embedding equation on $\Sigma_s$. Moreover,
\[
\lim_{s \to 0}  \Vert\tau_s  \Vert_{C^{5,\alpha}} = 0.
\]
\end{theorem}
\begin{proof}
The requirements for $|H|$ and $\alpha_H$ are clear from the proof of Theorem \ref{optimal-existence}. For the metric, $C^{5,\alpha}$ convergence guarantees the uniformness of the constants from the isometric embedding (see \cite[Section 4]{M-T-X} and \cite{Nirenberg}).
\end{proof}

\cite[Theorem 1]{C-W-Y2} immediately implies that the solution $\tau_s$ is a local minimum of the Wang--Yau quasi-local energy. See Appendix \ref{secB} for further discussions on the minimizing properties of the solution $\tau_s$ and the uniqueness of the optimal embedding. Next, we prove that the quasi-local energy with respect to the solution $ \tau_s$ converges to the quasi-local mass on the horizon as $s$ approaches $0$.
\begin{theorem} \label{limit_of_critical_point}
With the same assumption on the family of data over $\Sigma_s$ as in Theorem \ref{optimal-existence-family}, we have
\[
\lim_{s \to 0 } E(\Sigma_s, \tau_s) = m_{WY}(\Sigma_o)
\]
where $H_0$ is the mean curvature of $\Sigma_o$ in $\R^3$.
\end{theorem}
\begin{proof}
It is straightforward to see that the reference energy converges to
\[
m_{WY}(\Sigma_o)
\]
since the induced metric $\sigma_s$ converges in $C^{5,\alpha}$ and 
\[
\lim_{s \to 0}\Vert\tau _s \Vert_{C^{5,\alpha}} =0.
\]
As a result, it suffices to show that the physical Hamiltonian converges to $0$. Except for
\[
\int \theta_s \Delta_s \tau_s
\]
the other terms in the physical Hamiltonian converge to zero because $\tau$ and $|H|$ converge to zero. To rewrite this term, we consider the optimal embedding equation
\[
-(\widehat{H_s}\hat{\sigma}^{ab} -\hat{\sigma}^{ac} \hat{\sigma}^{bd} \hat{h_s}_{cd})\frac{\nabla_b\nabla_a \tau_s}{\sqrt{1+|\nabla\tau_s|^2}}+ div_{\sigma_s} (\frac{\nabla\tau_s}{\sqrt{1+|\nabla\tau_s|^2}} \cosh\theta_s|{H_s}|-\nabla\theta_s-\alpha_{H_s})=0
\]
 for $\tau_s$. Multiply the equation with  $\tau_s$ and integrate by parts. This gives us
\[
\begin{split}
  &- \int \theta_s \Delta_s \tau_s  \\
=& \int \tau_s  (\widehat{H_s}\hat{\sigma}^{ab} -\hat{\sigma}^{ac} \hat{\sigma}^{bd} \hat{h_s}_{cd})\frac{\nabla_b\nabla_a \tau_s}{\sqrt{1+|\nabla\tau_s|^2}} +  \frac{|\nabla \tau_s|^2}{\sqrt{1+|\nabla \tau_s|^2}} \cosh \theta_s |H_s| + \tau_s div_{\sigma_s} \alpha_{H_s}
\end{split}
\]
where the second line converges to zero.
\end{proof}

\section{Continuity of Wang--Yau quasi-local mass in the near horizon limit}\label{sec5}
In this section, we prove the continuity of the quasi-local mass in the near horizon limit. The following lemma about total mean curvature for surfaces in $\R^3$ will be useful:

\begin{lemma} \label{min-total-mean}
Let $(\Sigma,\sigma)$ be any surface with positive Gauss curvature. Let $H_0$ be the mean curvature of the isometric embedding of the metric $\sigma$ into $\R^3$. For any admissible time function $\tau$, let $\widehat \Sigma$ be the image of the isometric embedding of the metric $\hat \sigma= \sigma + d \tau \otimes d \tau$ and $ \widehat H$ be its mean curvature. We have
\[
\int_{\widehat \Sigma} \widehat H d \widehat \Sigma \ge \int_{\Sigma} H_0 d\Sigma.
\]
\end{lemma}
\begin{proof}
Let $K$ be the Gauss curvature of the metric $\sigma$ on $\Sigma$. We have $K>0$ by assumption. Moreover, since $\tau$ is admissible, we have, by, \cite[Definitioin 5.1] {W-Y}
\[
K + \frac{det (\nabla^2 \tau)}{1+ |\nabla \tau|^2}>0.
\]
It follows that for any $0 \le t \le 1$, the Gauss curvature $K(t)$ of the metric $\hat \sigma(t)=\sigma + d (t\tau) \otimes d (t\tau)$ is positive \footnote{This fact is proved and used already in \cite[Theorem 3]{C-W-Y2}. We include the proof here for completeness and to make it clear that the symmetry assumption in  \cite[Theorem 3]{C-W-Y2} is not needed for this elementary argument.} since
\[
K(t) = \frac{1}{1+t^2|\nabla \tau|^2} \left ( K + \frac{t^2 det (\nabla^2 \tau)}{1+t^2 |\nabla \tau|^2 } \right)
\]
As a result, $K(t)$ is clearly positive for all $t$ if $det (\nabla^2 \tau) \ge 0$. If $det (\nabla^2 \tau) < 0$, then $k(t)$ is a decreasing function in $t$ but we already know that $K(1) >0$.

Hence, the assumption of the Lemma implies that $\hat \sigma(t)$ admits a unique convex isometric embedding into $\R^3$. Let 
 $\widehat \Sigma(t) $ be the surface in $\R^3$  with the induced metric $\hat \sigma(t)$ and  $F(t)$ be the total mean curvature of the metric $\hat \Sigma(t)$ in $\R^3$. Namely
 \[
 F(t) =  \int_{ \widehat \Sigma(t) }\widehat H (t)d \widehat \Sigma(t) 
 \]
 where  $\widehat H(t)$ is the mean curvature of $\widehat \Sigma (t) $ in $\R^3$.
 
From \cite[Proposition 6.1]{W-Y} for the variation of total mean curvature, we get
\[
F'(t)   =  t \int_{ \widehat \Sigma(t) } \left (\widehat H (t) \hat \sigma^{ab} (t) - \hat h^{ab}(t) \right )  \tau_a \tau_b  d \widehat \Sigma(t) \ge 0
\]
where $\widehat H(t)$ and $\hat h(t)$ are the mean curvature and second fundamental of $\widehat \Sigma (t) $ in $\R^3$.

As a result,
\[
\int_{\widehat \Sigma} \widehat H d \widehat \Sigma = F(1) \ge F(0) = \int_{\Sigma} H_0 d \Sigma.
\]
\end{proof}
We are ready to prove the continuity of the Wang--Yau quasi-local mass in the near horizon limit. Namely, we prove that
\begin{equation} \label{WY_continuity}
\lim_{s \to 0 } m_{WY} (\Sigma_s ) = m_{WY} (\Sigma_o ).
\end{equation}

We look at the physical Hamiltonian and the reference Hamiltonian separately. Lemma \ref{min-total-mean} states that the minimum of the reference Hamiltonian is achieved at $\tau=0$. In the following, we find an upper bound of the physical Hamiltonian which converges to $0$ as the surface approaches the horizon. We first prove the following lemma for a surface with fixed induced metric $\sigma$ and connection 1-form $\alpha_H$

\begin{lemma} \label{physical_upper_lemma_1}
On a surface $\Sigma$ with fixed induced metric $\sigma$ and connection 1-form $\alpha_H$. There exists $\epsilon$ such that if
\[
|H| < \epsilon,
\]
then there exists $b>0$ such that for any time function $\tau$, we have
\[
\int_\Sigma \left[\sqrt{1+|\nabla\tau|^2}\cosh\theta|{H}| - \nabla \tau \cdot \nabla \theta -\alpha_H ( \nabla \tau) \right]d\Sigma < \cosh b \int_\Sigma |H| d\Sigma
\]
\end{lemma}
\begin{proof}
In this proof, all the integrals are integrated on $\Sigma$ with respect to the induced metric. Recall that from \cite[Proposition 2.1]{W-Y}, we have
\[
\begin{split}
     & \int \sqrt{(1+|\nabla \tau|^2)|H|^2+ (\Delta \tau)^2} - \Delta \tau \sinh^{-1} \frac{\Delta \tau}{|H|\sqrt{1+|\nabla \tau|^2}} \\
\le & \int ( \cosh \phi) \sqrt{(1+|\nabla \tau|^2)} |H|  + \phi \Delta \tau
\end{split}
\]
for any function $\phi$ on $\Sigma$. In particular, for any $b >0$, we have
\[
\begin{split}
     &  \int \sqrt{(1+|\nabla \tau|^2)|H|^2+ (\Delta \tau)^2} - \Delta \tau \sinh^{-1} \frac{\Delta \tau}{|H|\sqrt{1+|\nabla \tau|^2}}  \\
\le  & \int ( \cosh b) \sqrt{(1+|\nabla \tau|^2)} |H|  - b | \Delta \tau |
\end{split}
\]
by choosing $\phi = -b \sgn(\Delta\tau)$.

As a result, we have the following upper bound on the physical Hamiltonian (using $|H|<\epsilon$)
\begin{equation}
\label{physical_upper_form1}
\cosh b \int |H| + \epsilon \cosh b \int |\nabla \tau|   - b || \Delta \tau ||_{L^1} + || div \alpha_{H} ||_{L^{\infty}} || \tau ||_{L^1}
\end{equation}
Recall that we have the following inequality on the Sobolev norm (see, for example, \cite[Lemma 14]{Martinazzi}\footnote{While the result of Martizazzi is for domain with Dirichelet boundary condition, the same proof works for functions on $S^2$ with vanishing average.}): there exists a uniform constant $C$ such that
\[
C \Vert \Delta \tau \Vert_{L^1} > \Vert\tau\Vert_{W^{1,1}}.
\]
We first choose $b$ sufficient large such that $b \Vert \Delta \tau \Vert_{L^1}$ dominates $\Vert div \alpha_H \Vert_{L^{\infty}} \Vert \tau \Vert_{L^1}$ and then choose $\epsilon$ small such that $\epsilon \cosh b  \Vert \nabla \tau \Vert_{L^1}$ is also dominated by $b \Vert \Delta \tau \Vert_{L^1}$. This proves that the physical Hamiltonian is uniformly bounded from above by
\[
\cosh b \int |H|
\]
for some fixed $b$ when $\epsilon$ is sufficiently small.
\end{proof}
We apply Lemma \ref{physical_upper_lemma_1} to the family of surfaces $\Sigma_s$ and conclude that the physical Hamiltonian is uniformly bounded from above by
\[
\cosh b \int_{\Sigma_s} |H_s| d\Sigma_s.
\]
Similar to the proof of Theorem \ref{optimal-existence-family}, we have to make sure that the constants are controlled uniformly along the family. However, this requires much less restriction on the data compared to Theorem \ref{optimal-existence-family}. For example, it suffices to assume that $\sigma_s$ converges to $\sigma_o$ in $C^{4,\alpha}$ (This is to also make sure that the total mean curvature of the isometric embedding converges for the reference Hamiltonian), $|H|$ converges to $0$ in $L^{\infty}$ and $div \alpha_H$ is uniformly bounded in $L^{\infty}$

Combining the lower bound of the reference Hamiltonian and the upper bound of the physical Hamiltonian, we conclude that
\[
\lim_{s \to 0 } m_{WY} (\Sigma_s ) \ge m_{WY} (\Sigma_o ).
\]
Inequity in the other direction is straightforward, and we conclude that
\[
\lim_{s \to 0 } m_{WY} (\Sigma_s ) = m_{WY} (\Sigma_o ).
\]

\begin{theorem}
Assume that the horizon $(\Sigma_o,\sigma_o)$ satisfies the condition for Lemma \ref{mass_on_horizon}. Suppose we have a family of surfaces $\Sigma_s$ approaching the horizon $\Sigma_o$ such that the family of data $(\sigma_s,|H_s|,\alpha_{H_s})$ satisfies the following: $\sigma_s$ converges to  $\sigma_o$ in $C^{4,\alpha}$, $|H_s|$ converges to $0$ in $L^{\infty}$ and $div_{\sigma_s} \alpha_{H_s}$ is uniformly bounded in $L^{\infty}$. Then we have
\[
\lim_{s \to 0 } m_{WY} (\Sigma_s ) = m_{WY} (\Sigma_o ).
\]

\end{theorem}

\begin{remark}
We conjecture that 
\[
m_{WY} (\Sigma_s ) = E(\Sigma_s,\tau_s).
\]
Namely, the optimal embedding from Theorem \ref{optimal-existence} minimizes the Wang--Yau quasi-local energy for $s$ sufficiently small. 
\end{remark}

\appendix

\section{Leading order term of the solution to the optimal embedding equation}\label{secA}
In Section \ref{sec4}, we assume only that $|H|$ becomes small as $s$ approaches zero and that $\alpha_H$ is uniformly bounded. This assumption allows us to solve the optimal embedding equation for each sufficiently small $s$ near the horizon. In this section, we assume stronger conditions that
\[
\begin{split}
\lim_{s \to 0 } div_{\sigma_s} \alpha_{H_s} = & V \\
\lim_{s \to 0 } \frac{|H_s|}{s} =& h >0.
\end{split}
\]
Setting $\tau=0$ in \eqref{iteration-1}, we conclude from \eqref{iteration-1} and  \eqref{iteration-2} that
\[
F(0) = s \tau^{(1)} + o(s)
\]
where $ \tau^{(1)}$ is the solution to the linearized optimal embedding equation.
\begin{equation}
\Delta \sinh^{-1} \frac{\Delta \tau^{(1)}}{h} =V
\end{equation}
where $\Delta$ denote the Laplace operator on the horizon $\Sigma_o$. This is precisely \cite[Equation (6.4)]{Chen} \footnote{The last term on the right-hand side of \cite[Equation (6.4)]{Chen} vanishes from a previous discussion.}. We solve the equation as follows: We observe that
\[
\int_{\Sigma_o} V =0
\]
Let $v$ be any solution to
\[
\Delta v = V.
\]
There exists a unique constant $c$ such that
\[
\Delta \tau^{(1)} = h \left ( \sinh (v+c) \right)
\]
is solvable since $\sinh$ is monotone and $h>0$. 

From the contraction mapping, we conclude that
\[
\tau_s= s \tau^{(1)} + o(s).
\]
This verifies the conjecture in \cite[Chapter 6]{Chen} where an interactive scheme is proposed to find optimal embeddings on a family of surfaces approaching the apparent horizon. 
\section{Convexity of the Wang--Yau quasi-local energy}\label{secB}
In this section, we investigate two related problems: the uniqueness of the solution to the optimal embedding equation and the minimizing properties of the solution $\tau_s$ we found in Section \ref{sec4}.

Recall that from \cite[Theorem 1]{C-W-Y2}, a solution $\tau$ of the optimal embedding equation is a local minimum of the Wang--Yau quasi-local energy provided that 
\[
|H_{\tau}| > |H|>0
\]
where $H$ is the mean curvature vector of the surface in the spacetime $N$ and  $H_{\tau}$ is the mean curvature vector of the isometric embedding with time function $\tau$. 
\begin{lemma}
The solution $\tau_s$ from Theorem \ref{optimal-existence} is a local minimum for the Wang--Yau quasi-local energy on $\Sigma_s$ for $s$ sufficiently small.
\end{lemma}
\begin{proof}
By Theorem \ref{optimal-existence}, $\tau_s$ converges to $0$ in $C^{4,\alpha}$ as $s$ goes to $0$. Moreover, the Gauss curvature of $\Sigma_s$ is bounded by a positive constant from below. As a result, we obtain a uniform positive lower bound on $|H_{\tau_s}|$ (the norm of the mean curvature vector of the image of the isometric embedding of $\Sigma_s$ into $\R^{3,1}$ with time function $\tau_s$) for $s$ sufficiently small. On the other hand, $|H_s|$ goes to $0$ as $s$ goes to $0$. We conclude that 
\[
|H_{\tau_s}| > |H_s|>0
\]
for sufficiently small $s$. The lemma on follows from  \cite[Theorem 1]{C-W-Y2}.
\end{proof}
Following the proof of  \cite[Theorem 1]{C-W-Y2}, we prove the following lemma concerning the convexity of the Wang--Yau quasi-local energy.

\begin{lemma} \label{convexity_WY}
Let $\tau_0$ be an admissible time function on $(\Sigma,\sigma,\alpha_H)$. The Wang--Yau quasi-local energy $E(\Sigma,\tau)$ is convex at $\tau=\tau_0$ provided that 
\[
|H_{\tau_0}| > |H|>0
\]
where $H_{\tau_0}$ is the mean curvature vector of the isometric embedding of $(\Sigma,\sigma)$ with time function $\tau_0$. 
\end{lemma}

\begin{proof}
To compute the second variation of $E(\Sigma,\tau)$ at $\tau=\tau_0$, let $\Sigma_{\tau_0}$ be the image of isometric embedding into the Minkowski space with time function $\tau_0$ and write
\[
E(\Sigma,\tau) = E(\Sigma_{\tau_0},\tau) + \left ( E(\Sigma,\tau) - E(\Sigma_{\tau_0},\tau) \right)
\]
As in \cite[Theorem 1]{C-W-Y2}, $ E(\Sigma_{\tau_0},\tau)$ is convex at $\tau=\tau_0$ by positivity of quasi-local energy and 
\[
 E(\Sigma_{\tau_0},\tau_0)= 0
\] 

Moreover, $E(\Sigma,\tau) - E(\Sigma_{\tau_0},\tau) $ is
convex as long as $|H_{\tau_0}|> |H|>0$ ( \cite[Theorem 1]{C-W-Y2} shows that convexity for critical points. However, one may simply modify the linear term to apply the result in our situation).
\end{proof}

Fix any $\delta>0$, we consider the collection of time function $\tau$ such that 
\[
|H_{\tau}| > \delta
\]
for all $s$ sufficiently small. and let $\mathfrak C_{\delta}$ be the convex null of $\tau=0$ in this collection of time function. For all $s$ sufficiently small, the critical point $\tau_s$ belongs to $\mathfrak C_{\delta}$. Moreover, the Wang--Yau quasi-local energy $E(\Sigma_s,\tau)$ is convex  on $\mathfrak C_{\delta}$. This implies the following uniqueness and minimizing property of the solution $\tau_s$.
\begin{theorem}
For all $s$ sufficiently small, $\tau_s$ is the unique solution to the optimal embedding equation for the surface $\Sigma_s$ on the set $\mathfrak C_{\delta}$. Moreover, $\tau_s$ minimizes the quasi-local energy among all time functions contained in  $\mathfrak C_{\delta}$.
\end{theorem}

Instead of the convex hull, one can improve the above theorem by considering the intersection of star-shaped regions centered at $\tau_s$ within the subset of time function satisfying 
\[
|H_{\tau}| > \delta.
\]
Nevertheless, even after such improvement, one may still miss a large portion of the set of admissible time functions since we do not have a result for the mean curvature of isometric embedding analogous to Lemma \ref{min-total-mean} for the Gauss curvature.

\end{document}